MDPI

*Review*

# A Survey of Forex and Stock Price Prediction Using Deep Learning

Zexin Hu¹^ , Yiqi Zhao²^ and Matloob Khushi³,*

^ authors contributed equally both authors should be considered first-author

1   Affiliation 1; zehu4485@uni.sydney.edu.au
2   Affiliation 2; yzha9512@uni.sydney.edu.au
2   Affiliation 3; mkhushi@uni.sydney.edu.au
*   Correspondence mkhushi@uni.sydney.edu.au;

**Abstract:** The prediction of stock and foreign exchange (Forex) had always been a hot and profitable area of study. Deep learning application had proven to yields better accuracy and return in the field of financial prediction and forecasting. In this survey we selected papers from the DBLP database for comparison and analysis. We classified papers according to different deep learning methods, which included: Convolutional neural network (CNN), Long Short-Term Memory (LSTM), Deep neural network (DNN), Recurrent Neural Network (RNN), Reinforcement Learning, and other deep learning methods such as HAN, NLP, and Wavenet. Furthermore, this paper reviewed the dataset, variable, model, and results of each article. The survey presented the results through the most used performance metrics: RMSE, MAPE, MAE, MSE, accuracy, Sharpe ratio, and return rate. We identified that recent models that combined LSTM with other methods, for example, DNN, are widely researched. Reinforcement learning and other deep learning method yielded great returns and performances. We conclude that in recent years the trend of using deep-learning based method for financial modeling is exponentially rising.

## 1. Introduction

The share market is a snapshot of future growth expectation of the companies as well as the economy. Many factors have attributed to the stock's price fluctuation, which includes but not limited to macro-economic factors, the market anticipation and confidence in the company's management and operation. The advancement of technology allows the public to access a larger quantity of information in a timelier manner. It means that stock analysis has become more and more difficult as a considerable amount of data has to be processed in a relatively short time. People hope that the progress made in big data, especially in the Deep Learning field, can help them analyses stock information [1].

Forex is one of the largest financial markets in the world. The prediction of the exchange rate can provide investors with useful decision-making references to increase return and reduce risk. However, the exchange rate is always under the influence of many factors, such as countries' economy, politics, society, international situation, etc., so the complexity of the matter has made Forex prediction and forecasting a challenging research topic.[2] Nowadays, Forex Forecasting tasks apply many different deep learning models as the computer, and artificial intelligence technology matures.

Forex and stock are similar in many aspects. For example, they both have comparable technical indicators, both have similar charts (candle chart), and they would both be affected by its country's market sentiment. Therefore, this paper will discuss the application of deep learning to both Forex and the stock market and explore the impact of different deep learning methods on their price trend prediction accuracy.

The continuous development in the AI field leads to the wide use of deep learning techniques in many research fields and practical scenarios. Applications include natural language processing, image recognition, medical predictions, and more. The neural networks used in these applications have also developed and improved due to the rise of deep learning. For example, reinforcement learning has gained popularity since AlphaGo defeated the best chess player at the time using it, and reinforcement learning has been implemented in the financial prediction field since then [76]. These technological breakthroughs have given the stock and Forex prediction models a solid foundation to start and a greater room to improve.

The highly complex nonlinear relationship of deep learning can fully describe the complex characteristics of the influencing factors. Besides, many other fields have verified the accuracy of a deep learning model for prediction accuracy, such as image classification, gene analysis. Research results are also obtained for Time-series data analysis and prediction with a deep learning algorithm; for example, deep learning is used to predict the offline store traffic [3]. Overall, deep learning models have excellent performance in other research fields. Therefore, it is feasible to predict stock and Forex trends with deep learning.

Financial researchers around the world have been studying and analysing the changes in the stock and Forex market. The broadening application of artificial intelligence has led to an increasing number of investors using deep learning model to predict and study the stock and Forex price. It has been proven that the fluctuation of stock and Forex price could be predicted [4]. Different from the traditional statistical and econometric models, deep learning can describe complex influencing factors.

Therefore, this paper will investigate the different effects of different deep learning methods on stock and Forex forecasting according to the existing published papers. This survey will analyse each paper from the following aspects: 1. What is the dataset of this paper; 2.What is the variable of this paper;3. What kind of deep learning model had been adopted; 4. What is the result of the prediction model?

The structure of this paper will be as follows: firstly, the introduction of Forex and stock combined with deep learning; secondly, the criteria and research methods of the article selected by the survey, thirdly, the impact and analysis of different deep learning methods on stocks and Forex prediction; fourthly, the discussion and analysis of the above methods; finally, the conclusion of the whole paper.

## 2 Related deep learning methods and input introduction

### 2.1 Convolutional neural network (CNN)

CNN was widely used in the field of image recognition because of its powerful pattern recognition ability; its use was also extended to the field of economic prediction. Similar to the traditional neural network, CNN was composed of multiple neurons connected by a hierarchical structure, and the weights and bias between layers can be trained. CNN was different from the network structure of Fully Connected network such as DBN/SAE/BP, as the CNN could share the weight among the neurons in each layer of the network. Hence the model significantly reduced the weight of the network and avoided falling into dimensional disaster and local minimisation [5].

If the characteristics of the stock market at a specific time point were regarded as a feature graph, CNN had the potential to extract the characteristics of the stock market at the corresponding period from these feature graphs. Therefore, CNN could be used to build a timing-selection model and ultimately used to complete the construction of the timing-selection strategy.

## 2.2 Recurrent Neural Network (RNN)

RNN belonged to the neural network, and it was good at modelling and processing sequential data. The specific expression was that the RNN was able to memorise the previous state, and the previous state could be used in the current state calculation. The different hidden layers were non-independent, and the input of the current hidden layer included not only the output of the input layer but also the output of the previously hidden layer. For this reason, RNN would have a good performance in dealing with the sequential data.

The advantage of RNN was that it would consider the context of data in the process of training, which is very suitable for the scenario of stocks and Forex because the fluctuation at a particular time often contains some connection with the previous trend.

## 2.1.3 Long Short-Term Memory (LSTM)

LSTM model is one of the variants of the RNN. Its core contribution was to introduce the design of self-loop to generate the path of the gradient, which could continuously flow for an extended period. The weight of the self-loop was also updated in each iteration, which solved the gradient vanishing problem that was easily generated when the RNN model updated the weights [6].

The modelling of time series was essentially a process of nonlinear parameter fitting. The LSTM model could perform well to reveal the correlation of nonlinear time series in the delay state space, and to realise the purpose of stock prediction [7]. In the stock or Forex trend prediction model based on LSTM, it obtained the corresponding data characteristics from the stock or Forex history data.

## 2.4 Deep neural network (DNN)

DNN was a neural network with at least one hidden layer. It could provide modelling for complicated nonlinear functions and own a high-level abstraction ability which meant the fitting power of the model would be significantly improved. Meanwhile, it was a kind of discriminant model, which could be trained through the backpropagation algorithm.

Since the DNN was good at dealing with prediction problems with sizable data and complicated nonlinear mapping relations, an intelligent stock and Forex prediction system can be designed based on a DNN to predict stock and Forex trend. Hopefully, the model was able to achieve far higher accuracy than human beings.

## 2.5 Reinforcement learning

Reinforcement learning was one of the deep learning methods that focus on how to act according to the current situation to profit maximisation. In reinforcement learning, there were two basic elements: state and action. A strategy was defined as performing a particular action in a specific state. All the learner had to do was to learn a good strategy by continually exploring and learning.

If the state is regarded as the attribute and the action as the label, it was easy to know that both supervised learning and reinforcement learning was trying to find a map and inferred the label/action from the known attribute/state. In this way, the strategy in reinforcement learning was equivalent to the classification/regression in supervised learning. However, in practical problems, reinforcement learning did not have such labelling information as supervised learning, and the results were often obtained after the attempt of the action. Therefore, reinforcement learning would continuously adjust the previous strategy through the feedback of the result information, for this reason, the algorithm would learn: in which state to take which step would have the most beneficial result.

Therefore, reinforcement learning could learn the best timing trading action (selecting the best price, trading duration, and order size) according to the market response. It can view the contextual information (price, news, public opinion, interest rate, fees, trading action, returns, losses, etc.) of the transaction as an environment of reinforcement learning. Gains or losses could be thought as the reward for learning, trading actions as actions, and factors as states or observations to realise the prediction of the stock and Forex trend.

## 2.5 Other deep learning methods

In this paper, we will discuss the prediction of stock and Forex trend by other deep learning methods, for example, Hybrid Attention Networks (HAN) and self-paced learning mechanism (NLP); multi-filters neural network (MFNN), and Wavenet. The frequency of these methods in the selected articles is not high so that they will be discussed in section 4.7.

## 3. Review methodology and criteria

### 3.1 Paper selection methods

In the past few years, there had been many deep learning model papers on stock and Forex forecasting. In this paper, the articles analysed in this paper were all from the DBLP computer science bibliography and Microsoft Academic.

Firstly, the keywords were searched in DBLP are:' CNN stock/Forex';' LSTM stock/Forex';' Deep learning stock/ Forex';' RNN stock/Forex' and' Reinforcement learning stock/Forex'. The keywords were searched in Microsoft Academic, then the filter of '2015-2021' and 'Deep learning' were applied.

Secondly, The quality of the selected articles from DBLP was ensured by excluding all the journals and conferences that were informally published. The quality of the selected articles from Microsoft Academic were controlled by excluding all the journals and conferences that were informally published as well as a minimum of 5 citations. Filtering on citations as the searching method was designed to detect under-study area with in this field, the paper with 5 more citations can be an indication that the area could be potentially explored. Furthermore, it should be noted that we only consider papers with a novel model, implementing existing model would not be analysised in this review.

At the same time, the timeliness of the survey was provided by focusing on publications after 2015. Among them, there were 4 papers in 2015,1 papers in 2016, 15 papers in 2017, 30 papers in 2018, 28 papers in 2019, and 10 papers in 2020, in total 88 existing papers would be reviewed.

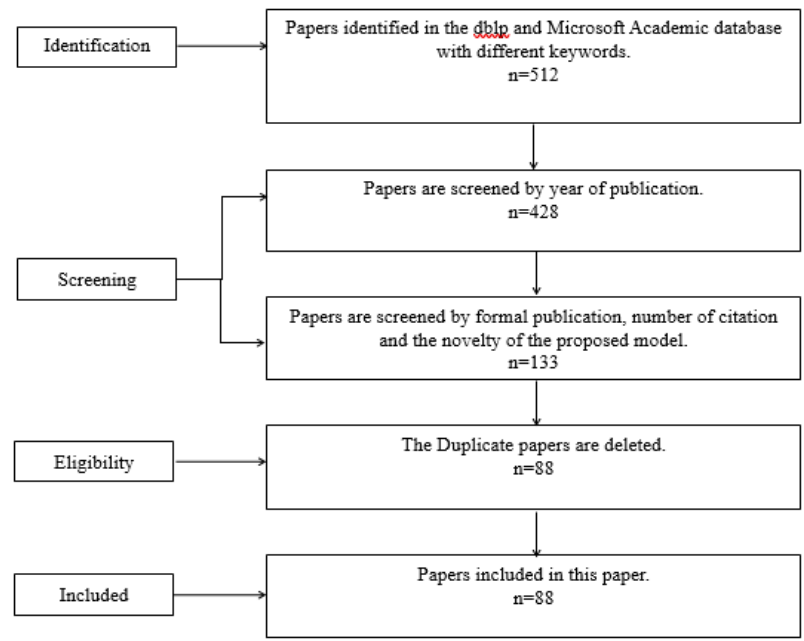

Figure 1 The systematic literature review in this paper

### 3.2 Selected paper statistics

In this paper, the goal was to study the use of six different deep learning methods in Forex/stock (which are CNN, LSTM, DNN, RNN, Reinforcement learning, and other deep learning methods) with different datasets, input variables, and model type. All the results were compared together for discussion and for concluding after individually analysed.

Figure 1 showed the annual distribution of papers collected and reviewed. Figure 2 showed the distribution of the different methods in this paper. Figure 3 showed the systematic literature review in this paper. Table 1 showed the distribution of different types of articles in this paper.

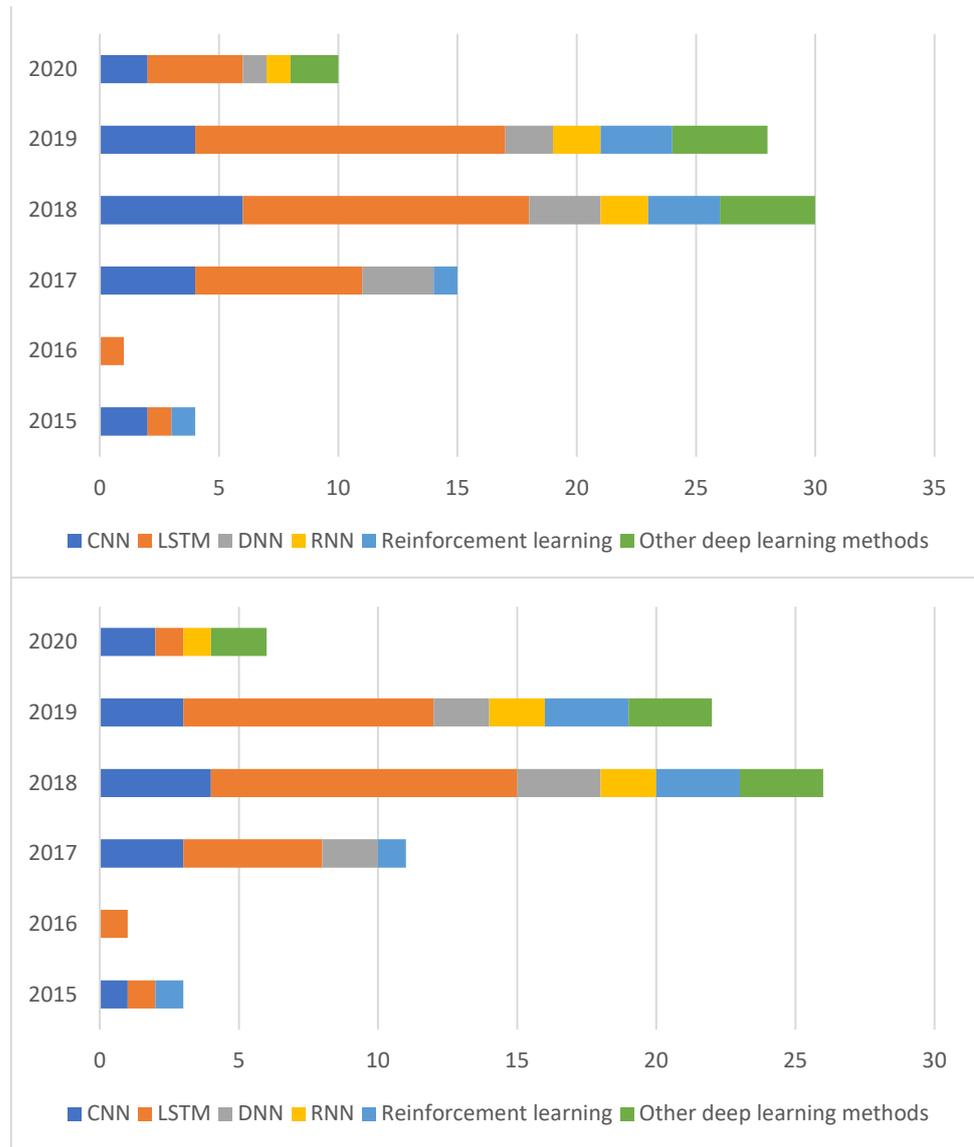

Figure 2 The annual distribution of papers collected and reviewed

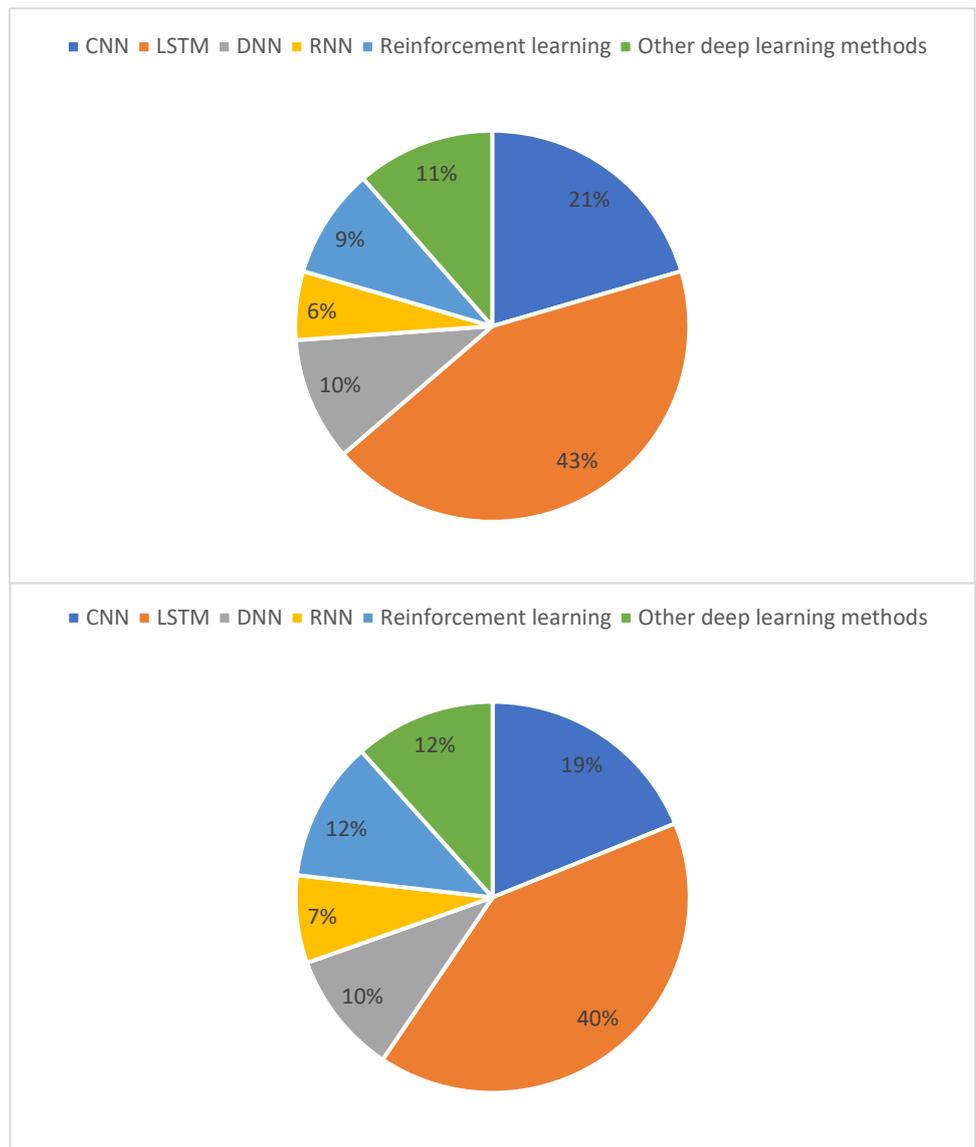

Figure 3 The distribution of the different methods in this paper

Table1. The distribution of different types of articles in this paper

| Type of Method | Total paper | Number of papers in journals | Number of papers in conferences |
| --- | --- | --- | --- |
| CNN | 18 | 6 | 12 |
| LSTM | 38 | 13 | 25 |
| DNN | 9 | 5 | 4 |
| RNN | 5 | 2 | 3 |
| Reinforcement Learning | 8 | 1 | 7 |
| Other deep learning methods | 10 | 3 | 7 |

## 4. Results

### *4.1 Papers Descriptions*

### 4.1.1 CNN

17 articles that used CNN technology for the stock and Forex prediction were briefly described below, and as shown in the table. Table 2 shows the author, variables, dataset, and model of all the papers mentioned.

Table 2 The overall information for the papers which used the CNN model

| Reference No. | Author | Dataset | Variables | Model |
|---|---|---|---|---|
| [8] | Maqsood, H. | 1.Top 4 performing companies of US, Hong Kong, Turkey, and Pakistan 2.Twitter dataset | 1.Open price, high price, low price, AdjClose price, Volume, and Close price 2.sentiment(Positive, neutral and negative sentiment). | CNN |
| [9] | Patil, P. | 1. News collected on a financial website. 2.30 stocks of fortune 500 companies, such as WMT, XOM, and AAPL | Adjacency matrix calculated by the correlation coefficient and using news co-mentions | Graph convolution neural network (GCN) |
| [10] | Sim, HS. | S&P 500 minute data | Close price | CNN |
| [11] | Hoseinzade, E. | S & P 500, NASDAQ, Dow Jones industrial average, NYSE, and Russell | 82 variables including high, low, close price; Volume, RSI, KD, WR.etc | 2D-CNN And 3D-CNN |
| [12] | Eapen, J. | Standard and Poor's (S&P) 500 stock dataset | Close price | CNN--Bi-Directional LSTM |
| [13] | Yang, H. | S&P 500 Index ETF (SPY) | high, low,close price;Volume,RSI,KD,WR,ROC,CCI | CNN with MICFS |
| [14] | Cai, S. | 1.Crawling financial news 2.Baidu Index | 1.word vector; headline and keyword training set in the news 2.Close price | CNN-LSTM |
| [15] | Oncharoen, P. | Reuters and Reddit Standard & Poor's 500 Index (S&P500) and Dow Jones Industrial Average (DJIA) | Word vectors of headlines Close prices, Bollinger band, RSI, and Stochastic Oscillator | CNN-LSTM |
| [16] | Liu, Y. | 1.Thomson Reuters 2.Standard & Poor's 500 index (S&P 500) | 1.Financial news corpus with headlines from Apple 2.. Open price; Close price; High price; Low price; Volume; Stochastic Oscillator (%K); Larry William (LW) %R indicator; and Relative Strength Index (RSI) | TransE-CNN-LSTM |
| [17] | Selvin, S. | NSE listed companies : Infosys, TCS, and CIPLA | Close price | CNN Sliding-window model |
| [18] | Liu, S. | Chinese stocks from the SINA FINANCE(Not given specific data) | Close price | CNN-LSTM |
| [19] | Gudelek, M.U. | Exchange-Traded Funds(ETF) | Close price, RSI, SMA, MACD, MFI, Williams %R, the stochastic oscillator, the ultimate oscillator | 2D-CNN |

| [20] | Ding, X. | 1.S&P 500 index 2.Reuters and Bloomberg | 1.close price 2.long-term, mid-term and short-term feature vectors of news headlines | NTN-CNN |
|---|---|---|---|---|
| [71] | Zhao, Y. and Khushi | USDJPY Exchange rate | 1. Momentum Indicators: RSI5, RSI10, RSI20, MACD, MACDhist, MACDsignal, Slowk, Slowd, Fastk, Fastd, WR5, WR10, WR20, ROC5, ROC10, ROC20, CCI5, CCI10, CCI20 2. Volume Indicators: ATR5, ATR10, ATR20 , NATR5, NATR10, NATR20, TRANGE | Wavelet Denoised-ResNet CNN with light GBM |
| [76] | Chen, S. and He, H. | Chinese stock market | Closing price | CNN |
| [77] | Ugur Gudelek, M. | SPDR ETF | Closing price and technical indicators | CNN |
| [78] | Chen, Jou-Fan | Taiwan index futures | Open, high, low, closing price | GAF+CNN |
| [90] | Wen, M | S%P500 | Open, High, Low, Close, Adjusted Close, and Volume. | CNN |

Maqsood, H. proposed a CNN model that made use of the price and sentiment analysis as input and compared the proposed model with the Linear Regression and SVM. He concluded that not all the significant events have a serious impact on stock exchange prediction. However, more important local events could affect the performance of prediction algorithms [8].

Patil, P. proposed a new network using graph theory and CNN, which leveraged Spatio-temporal relationship information between different stocks by modelling the stock market as a complex network. Meanwhile, the model used both stock indicators and financial news as input [9].

Sim, HS proposed a CNN network that using the 9 technical indicators (Close Price, SMA, EMA, ROC, MACD, Fast%K, Slow%D, Upper Band, Lower Band) to verify the applicability of the CNN method in the stock market. And he concluded that the use of technical indicators in stock price forecasting by CNN has no positive effect [10].

Hoseinzade, E. proposed two models: 2D-CNN and 3D-CNN using 82 different technical indicators. And These two structures could improve the predictive performance of the baseline algorithm by about 3% to 11% [11].

Eapen, J.proposed a model that had multiple pipelines of CNN and bi-directional LSTM units. And it could improve prediction performance by 9% using a single pipeline deep learning model and by over a factor of six using support vector machine regressor model on S&P 500 grand challenge dataset [12].

Yang, H.m, proposed a multi-indicator feature selection for stock index prediction based on a multichannel CNN structure without sub-sampling, named MI-CNN framework. In this method, candidate indicators were selected by the maximal information coefficient feature selection (MICFS) approach, to ensure the correlation with stock movements whilst reducing redundancy between different indicators [13].

Cai, S. proposed the CNN and LSTM forecasting system with financial news and historical data of the stock market. It had generated seven prediction models. According to the ensemble learning method, the seven models are constructed into one ensemble model to obtain an aggregated model. Unfortunately, all models had a lower prediction accuracy [14].

Oncharoen, P. proposed a new framework to train a DNN for stock market prediction. A new loss function was developed by adding a risk-reward function, which was derived from the trading simulation results [15].

Liu, Y. proposed to incorporate a joint model using the TransE model for representation learning and a CNN, which extracted features from financial news articles. And then he combined the model with LSTM to predict the price movement. The model could improve the accuracy of text feature extraction while reducing the sparseness of news headlines [16].

Selvin, S. used CNN, LSTM, and RNN architectures to forecast the price of NSE listed companies and compares their performance. The final results showed that CNN is the best architecture to predict the price of the stock because it could identify the trend of the directional change [17].

Liu, S.proposed a CNN-LSTM model, and the model performed basic Momentum strategy and Benchmark model whose return rates were 0.882 and 1.136 respectively. The CNN part could extract useful features even from low signal-to-noise time-series data, and the LSTM part could predict future stock prices with high accuracy, then the predicting outcomes were used as timing signals [18].

Gudelek, M.U. proposed a 2D-CNN model. This model adopted a sliding window approach then generated images by taking snapshots that are bounded by the window over a daily period. The model could predict the next day's prices with 72% accuracy and end up with 5:1 of our initial capital [19].

Ding, X. proposed a deep learning method for the event-driven stock market prediction. Firstly, events were extracted from news text and represented as dense vectors, trained using a novel neural tensor Network(NTN). Secondly, CNN was used to model both short-term and long-term influences of events on stock price movements [20].

Zhao, Y. proposed a Wavelet Denoised-ResNet CNN for Forex exchange rate prediction. The technical indicators were treated as an image matrix, the image matrix was first denoised using Wavelet method, then processed by ResNet CNN, and lastly, LightGBM was used to replace the softmax layer to output a prediction. [71]

Chen, S. and He, H. proposed a CNN model with a novel architecture, where it generated better results than benchmark RNN model. [76]

Ugur Gudelek, M proposed a CNN model where financial data is taken as a image, then using CNN to predict the desired trading action, the model yielded reasonable results. [77]

Chen, Jou-Fan proposed a novel CNN model which utilized the power of Gramian Angular Field, the results produced were average, but it is a interesting research direction. [78]

Wen, M proposed a CNN model which relied on the reconstructed of time series, it turned the time series into segment, then CNN was used to classify each segment. The model generated good results.[90]

### 4.1.2 RNN

5 articles that used RNN technology for the stock and Forex prediction were briefly described below, and as shown in the table. Table 3 showed the author, variables, dataset, and model of all the papers mentioned above.

Table 3 The overall information for the papers which used RNN model

| Reference no. | Author | Dataset | Variables | Model |
|---|---|---|---|---|
| [53] | Ni, L. | EURUSD,AU-DUSD,XAU-USD,GBPJPY,EUR-JPY,GBPUSD,USDCHF,USDJPY and USDCAD | open price, close price, highest price and lowest price | C-RNN |
| [54] | Li, C. | CSI300,CSI200and CSI500 | Open price, high price, low price, close price, Volume, and amount. | Multi-task RNN with MRFs |
| [55] | Chen, W | HS300 | 1.Technical features: open price, high price, low price, | RNN-Boost |

| | | | close price, Volume, price change, price limit, Volume change, Volume limit, Amplitude, and difference. 2.Content features: sentiment features and LDA features | |
| [56] | Zhang, R. | Sandstorm sector of Shanghai Stock Exchange | Open price, close price, highest price, lowest price, and the daily trading volume | C-RNN(DWNN) |
| [73] | Zeng, Z. and Khushi | USDJPY Exchange rate | Momentum indicators: average directional movement index, absolute price oscillator, arron oscillator, balance of power, commodity channel index, chande momentum oscillator, percentage price oscillator, moving average convergence divergence, Williams, momentum, relative strength index, stochastic oscillator, triple exponential average. • Volatility indicators: average true range, normalised average true range, true range. | Attention-based RNN-ARIMA |

Ni, L. proposed a CRNN model to predict the price of 9 kinds of forex pair. The results showed that the proposed model performed much better than the LSTM model and the CNN model[53].

Li, C. proposed a Multi-task RNN model with MRFs. The MMPL is proposed to automatically extract diversified and complementary features from individual stock price sequences which means there is no need for the technical indicators. Features learned by MMPL were passed to a binary MRF with a weighted lower linear envelope energy function to utilise intra-clique higher-order consistency between stocks[54].

Chen, W proposed an RNN-Boost model that made use of the technical indicators and sentiment features, and Latent Dirichlet allocation (LDA) features to predict the price of stocks. Its results showed that the proposed model outperformed the single-RNN model[55].

Zhang, R. proposed a Deep and Wide Neural Networks (DWNN) model where CNN's convolution layer was added to the RNN's hidden state transfer process. The results showed that the DWNN model could reduce the prediction mean squared error by 30% compared with the general RNN model[56].

Zeng. Z proposed a novel Attention-based RNN (ARNN) where wavelet denoised input was passed to ARNN. The forecast was calculated using the ARIMA and the output of ARNN model. [73]

### 4.1.3 LSTM

27 articles that used LSTM technology for the stock and Forex prediction were briefly described below, and as shown in the table. Table 4 showed the author, variables, dataset, and model of all the papers mentioned.

Table 4 The overall information for the papers which used LSTM model

| # | Author | Dataset | Variables | Model |
| --- | --- | --- | --- | --- |
| [21] | Nikou, M. | iShares MSCI United Kingdom | Close price | LSTM |
| [22] | Fazeli, A. | S&P 500 | Open price, high price, low price, close price, adjusted close price, Volume, volatility, | LSTM |

| | | | Williams %R and RSI | |
|---|---|---|---|---|
| [23] | Xu, Y. | Microsoft (MSFT)、PO logistics (XPO) and AMD Daily stock price data are collected from Yahoo Finance from 11 in-dustries Finance tweets from a social media com-pany StockTwits | Open price, high price, low price, close price,AD,ADX,EMA ,KAMA,MA,MACD, RSI,SAR,AMA .etc and finance tweet sentiment | Attention-based LSTM |
| [24] | Lakshminarayanan, S.K. | Dow Jones Industrial Average (DJIA) | Close price, moving average, Crud oil price, gold price | LSTM |
| [25] | Rana, M. | Spanish stock com-pany Acciona | Close price | LSTM |
| [26] | Naik, N. | CIPLA stock, ITC stock, TCS stock, ONGC stock, and Nifty in-dex | Close price | RNN with LSTM |
| [27] | Nguyen, D.H.D | NASDAQ stock mar-ket:GE,AAPL,SNP and FB | Trade volume, open, close, high, low, and adjusted close prices | Dynamic LSTM |
| [28] | Lai, C.Y. | Foxconn, Quanta, Formosa Taiwan Cement and Taiwan Semiconduc-tor | KD, OBV, MACD, RSI, and the average of the previous five days stock market in-formation (open, high, low, Volume, close) | LSTM |
| [29] | Hossain, M.A. | S&P500 | Open price, close price, volume | LSTM with GRU |
| [30] | Baek, Y. | KOSPI200 and S&P500 | Close price | LSTM with preven-tion module, predic-tion module |
| [31] | Kim, H.Y. | KOSPI 200 | Close price | GEW-LSTM |
| [32] | Li, H. | CSI-300 | Open price | Attention-based Multi-Input LSTM |
| [33] | Cheng, L.-C. | Data from Taiwan Stock Exchange Cor-poration((Not given specific data) | Open price, close price, low price, high price, Volume, KD, MA, RSV, etc. | Attention-based LSTM |
| [34] | Shah, D. | Tech Mahindra (NSE: TECHM) BSESensex | Close price | LSTM |
| [35] | Lin, B.-S. | Taiwan Stock Ex-change Corporation (TWSE) | Trade volume, trans-action, open price, highest price, lowest price, close price, KD, RSI, and Bol-linger Bands(BBands) | LSTM |
| [36] | Skehin, T. | Facebook Inc. (FB), Apple Inc. (AAPL), Amazon.com Inc (AMZN), Netflix Inc. (NFLX) and Alpha-bet Inc. (GOOG) in NASDAQ of S&P 500 | Close price | ARIMA-LSTM-Wavelet |
| [37] | Zhang, X. | China's A-share mar-ket | The open price, close price, highest price, lowest price, and trading volume and 11 indicators | RNN with LSTM |
| [38] | Achkar, R. | Facebook stocks, Google stocks, and Bitcoin | Close price | RNN with LSTM |

| | | stocks collected from Yahoo finance | | |
|---|---|---|---|---|
| [39] | Zeng, Y. | SSE50 index | N/A | LSTM |
| [40] | Shao, X. | Ping An Bank | Close price | Kmeans-LSTM |
| [41] | Zhao, Z. | SSE Composite Index,SSE 50,CSI 500,CSI 300 and SZSE Composite Index | Close price | Time-weighted LSTM |
| [42] | Nelson, D.M. | IBovespa index from the BM&F Bovespa stock exchange | Open price, close price, high price, low price and Volume exponentially weighted moving averages, etc.(175 indicators in total) | LSTM |
| [43] | dos Santos Pinheiro | 1.Standard&Poor's 500 index 2.Reuters and Bloomberg | 1.financial domain news text(headlines instead of the full content); 2.Close price | NLP+LSTM |
| [44] | Buczkowski, P. | Expert recommendation from TipRanks company | A stock identifier, a list of expert recommendations of varying length, and optional target labels (class) | LSTM with GRU |
| [3] | Akita, R. | Morning edition of the Nikkei newspaper Nikkei 225 | 1.Paragraph Vector; 2open, close, highest and lowest price | LSTM |
| [45] | Chen, K | China stock market in Shanghai and Shenzhen from Yahoo finance | volume, high, low, open, close price | LSTM |
| [75] | Qi, Ling and Khushi | Forex Exchange rate | Technical indicators | LSTM |
| [80] | Pang, Xiong Wen | Chinese stock | stock data, stock news, capital stock and shareholders, and financial analysis | DeepLSTM with encoder |
| [81] | Feng, Fuli, et al. | NASDAQ and NYSE | Stock data | LSTM with ranking relation |
| [82] | Chung, Hyejung, and Kyung-shik Shin. | Korea Stock Price Index | Open, High, low, closing, volume and technical indicators. | LSTM with GA |
| [83] | Li, Jiahong, et al. | Chinese Stock Market | Closing price and sentiment data | LSTM with Naïve Bayes |
| [84] | Zhang, Kang, et al. | S&P 500 Index, ShanghaiComposite Index, IBM from NYSE, MSFT from NASDAQ, PingAn Insurance Company of China (PAICC) | Open, High, low, closing, volume and technical indicators. | LSTM with Generative Adversarial Network (GAN) |
| [91] | Jin, Zhigang, et al. | Apple stock price | Sentiment and stock price data | LSTM with sentiment analysis model |
| [92] | Long, Jiawei, et al. | Chinese Stock market | Market information including transaction records | LSTM with CNN |
| [93] | Chen,M. | Chinese stock market | Sentiment information | LSTM |
| [94] | Qian,F. Chen,X | Chinese stock market | Closing price | LSTM with ARIMA |
| [95] | Li,Z. Tam,V. | Asian stock indexes | Closing price and technical indicators | LSTM with Wavelet Denosing |

Nikou, M. proposed an LSTM model and compared it with the ANN model, SVR model, and RF model. The results showed that the LSTM model performed better in the prediction of the close price of iShares MSCI United Kingdom than the other models mentioned in the paper[21].

Fazeli, A. proposed an LSTM model; his result showed an improve through using LeakyReLU. MSE came below 0.1. Whilst compared with the ReLU function; the MSE went well above 0.2. He examined the effect of RSI, Williams %R, and volatility on the loss of the model. It was shown that the model's loss was reduced by using only RSI, which contributed to the performance of the model[22].

Xu, Y. proposed an attention-based LSTM model that performs better than the regular LSTM. It was found that finance tweets that were posted from market closure to market open the next day had more predictive power on the next day's stock movement. The weighted sentiment on the max follower on StockTwits also performed much better than other methods[23].

Lakshminarayanan, S.K. proposed an LSTM model combined with Crud oil price, gold price, and moving average, which performed much better than the LSTM model without them and the SVM model. It showed that the Crud oil and gold price had some impact on the stock price prediction[24].

Rana, M. proposed an LSTM model that outperformed the LR and SVR model. He also compared the different activation functions with different optimisers and concluded that the tanh activation with the adam algorithm performs best with the accuracy of 98.49%[25].

Naik, N. proposed an RNN with the LSTM model. The model had the ability to keep the memory of historical stock returns in order to forecast future stock return output. It was worth noting that the recent stock information rather than old related stock information was stored and used. The network also outperformed the Feed Forward ANN model[26].

Lai, C.Y. proposed a Dynamic LSTM model The results showed that stock prediction accuracy based on MAE, MAPE, RMSE, and MSE obtained by the dynamic LSTM model was much better than that by the static LSTM model. The dynamic model also consistently outperformed the linear models SA-5 and EMA-0.5 when predicting four stocks[27].

Lai, C.Y. proposed an LSTM model which used average previous five days stock market information (open, high, low, Volume, close) as the input value. The initial prediction was calculated using the value. The prediction was then used as part of the average of the stock price information for the next five days through the ARIMA method. Moreover, he utilised Technical Analysis Indicators to consider whether to buy stocks or continue to hold stocks or sell stocks[28].

Hossain, M.A. proposed an LSTM model followed by GRU. Both of the LSTM and GRU were powerful recurrent networks that could perform better and faster in terms of accuracy in regression-based prediction. And the proposed model outperformed the LSTM only, GRU only, and GRU followed by the LSTM model[29].

Baek, Y. proposed a novel data augmentation approach for stock market index forecasting through ModAugNet framework, which consisted of two modules: an overfitting prevention LSTM module and a prediction LSTM module. The overfitting problems mainly caused by the limited availability of data points for training[30].

Kim, H.Y. proposed some LSTM model to forecast stock price volatility that combined the LSTM model with various generalised autoregressive conditional heteroscedasticity (GARCH) -type models. He found that the GEW-LSTM model, which combined all three models, GARCH, EGARCH, and EWMA, with LSTM, performed best[31].

Li, H. proposed an improved MI- LSTM based on LSTM and attention mechanism, which achieved better performance in extracting potential information and filtering noise. The model could assign different weights to different input series to keep the dominant status of the mainstream while absorbing information from leaky input gates[32].

Cheng, L.-C. proposed an Attention-based LSTM model that could solve the problem of exploding and vanishing gradients and thus did not effectively capture long-term dependencies[33].

Shah, D. proposed an LSTM model which was compared with the DNN model. The proposed model was able to predict volatile movements in the true data. In general, it was able to recognise the directionality of the changes in the true data more accurately than the DNN[34].

Lin, B.-S. proposed an LSTM model to predict the price of the top 10 industries included in Taiwan Stock Exchange Corporation (TWSE).In the experimental results, most of the results were reasonable except for the MTK stock[35].

Skehin, T. proposed a linear Autoregressive Integrated Moving Average (ARIMA) model and LSTM network for each series to produce next-day predictions. Wavelet methods decomposed a series into approximation and detail components to better explain behaviour over time. He combined these techniques in a novel ensemble model in an attempt to increase forecast accuracy[36].

Zhang, X. proposed a simple but efficient method to predict future stock return ranking without handcrafted features[37].

Achkar, R. proposed an approach to predict stock market ratios using artificial neural networks. It considered two different techniques BPA-MLP and LSTM-RNN their potential, and their limitations. And the LSTM-RNN model outperformed the other one slightly[38].

Zeng, Y. proposed an LSTM model which is based on the dataset SSE50, and the results showed that the accuracy is above 65%[39].

Shao, X. proposed a Kmeans-LSTM model that used the time window to divide the stock price sequential data into several equal sub-sequences. And the K-means algorithm is used to cluster the stock price sub-sequence (He did not give the specific result data.) [40].

Zhao, Z. proposed a time-weighted LSTM. Unlike other models which treat data indiscriminately, the proposed model could carefully assign weights to data according to their temporal nearness towards the data that was used for prediction. And the results showed that the proposed model outperformed the SVM, RNN, and Random Forest, model[41].

Nelson, D.M. proposed an LSTM model which was based on the price history and technical analysis indicators. And the results showed that the proposed model was better than the MLP and Random Forest model[42].

Dos Santos Pinheiro proposed a character-based neural language model for an event-based trading model that was combined with NLP and LSTM. The results showed that the proposed model performed better than some model proposed in other papers such as WI-RCNN and SI-RCNN[43].

Buczkowski, P. tackled the problem of stock market predicting feasibility, especially when predictions were based only on a subset of available information, namely: financial experts' recommendations. The analysis was based on data and results from the ISMIS 2017 Data Mining Competition[44].

Akita, R. proposed an approach that converted newspaper articles into their distributed representations via Paragraph Vector and modelled the temporal effects of past events on open prices about multiple companies with LSTM[3].

Chen, K proposed an LSTM model to predict the price of the Chinese stock. But the results showed that the accuracy is only 27.2%[45].

Qi, L. proposed an LSTM model to predict the price of the forex exchange rate. The model used technical indicators to calculate events then LSTM is used to make the prediction based on the events. [75]

Pang, Xiong Wen propsed two improved deep LSTM model with embedded layers. The results showed its improvement over the benchmark. [80]

Feng, Fuli, et al. proposed a noval LSTM model that combined with stock relational model, it had a great performance compared with its benchmarks. [81]

Chung, Hyejung, and Kyung-shik Shin proposed a LSTM model which is optimized by genetic algorithm (GA). The novel hybrid model had outperformed the benchmark.[82]

Li, Jiahong, et al. proposed a LSTM model that was combined with a Naïve Bayne's model. This hybrid model had outperformed benchmarks, and delivered promising results. [83]

Zhang, Kang, et al. propsed a novel generative adversarial network based on MLP and LSTM. The model is able to better results than some of the vanilla models. [84]

Jin, Zhigang, et al. propsed to incorporate sentiment analysis model into LSTM, they successfully created a novel model which deliver a reasonable result. [91]

Long, Jiawei, et al. propsed a hybrid model, which utilised CNN to extract transactional information, then it was passed through a LSTM to predict the stock price. The model showed improvement over some of the vanilla benchmarks.[92]

Chen,M proposed a LSTM model which uses the sentiment data collected. It combined both sentimental model as well as LSTM to produce a good result. [93]

Qian,Fei. Proposed a LSTM model which used ARIMA to improve its prediction power. The model yielded a reasonable result.[94]

Li.Z and Tam, V proposed a LSTM model that used wavelet denoising technique before passing through the LSTM model, the model had produced good result. [95]

### 4.1.4 DNN

13 articles that used DNN technology for the stock and Forex prediction were briefly described below, and as shown in the table. Table 5 shows the author, variables, dataset, and model of all the papers mentioned.

Table 6 The overall information for the papers which used DNN model

| # | Author | Dataset | Variables | Model |
|---|--------|---------|-----------|-------|
| [46] | Song, Y. | KOSPI and KOSDAQ | 715 novel input-features(such as (moving average and disparity of stock price) | DNN |
| [47] | Naik, N. | NSE ICICI Bank SBI Bank Kotak Bank Yes Bank | SMA, Exponential moving average, Momentum indicator, Stochastic oscillator, Moving average convergence divergence, Relative strength index, Williams R, Accumulation Distribution index, Commodity channel index | DNN |
| [48] | Chatzis, S.P. | FRED database and the SNL | Stock price, Exchange rates，VIX index，Gold price, TED spread, Oil price, Effective Federal Funds Rate, High yield bond returns | DNN |
| [49] | Abe, M. | MSCI | 25 variables:1.Book-to-market ratio 2.Earnings-to-price ratio3.Dividend yield4.Sales-to-price ratio5.Cash flow-to-price ratio 6.Return on equity 7.Return on asset 8.Return on invested capital 9.Accruals10.Sales-to-total assets ratio 11.Current ratio12.Equity ratio13.Total asset | DNN |

| | | | growth14.Investment growth15.Invest-ment-to-assets ratio 16.EPS Revision(1 month) 17.EPS Revision(3 months) 18.Market beta19. Market value 20. Past stock return(1 month) 21.Past stock return(12 months) 22 .Volatil-ity23.Skewness24.Id-iosyncratic volatil-ity25. Trading turno-ver | |
|---|---|---|---|---|
| [50] | Nakagawa, K. | TOPIX index | 16varia-bles:60VOL,BETA,S KEW,ROE,ROA,AC CRUALS,LEVER-AGE,12-1MOM,1MOM,60M OM,PSR,PER,PBR,P CFR,CAP,ILIQ | Deep factor model(DNN) with layer-wise relevance propagation and mul-tiple factors) |
| [51] | Chong, E. | KOSPI | Close price | DNN with autoen-coder |
| [52] | Singh, R. | NASDAQ | 36variables:Open price,high price, low price, close price, MA5,MA10,MA20, BIAS5,BIAS10,DIF F,BU,BL,K,D,ROC, TR,MTM6,MTM12, WR%10,WR%5,OS C6,OSC12,RSI6,RSI 12,PSY and the deri-vation of their calcu-lation | (2D² PCA) +DNN |
| [87] | Yu, Pengfei, and Xuesong Yan | S&P 500, DJIA, the Nikkei 225 (N 225), the Hang Seng index (HSI), the China Se-curities index 300 (CSI 300) and the ChiNext index | Closing price | DNN with phase-space reconstruction (PSR) and LSTM |
| [89] | Yong, Bang Xiang, et al. | Singapore stock mar-ket | Intraday prices | DNN |

Song, Y. proposed a DNN with 715 novel input-features configured on the basis of technical analysis. He also had proposed a plunge filtering technique to improve the accuracy of the training model by collecting similar stocks. It was worth noting that the proposed model had great profitability[46].

Naik, N. proposed a DNN model that used the Boruta feature selection technique to solve the problem of technical indicator feature selection and identification of the relevant technical indica-tors. And the results showed that his model performed much better than ANN and SVM model[47].

Chatzis, S.P. proposed a DNN model in which it used Boosted approaches to predict stock market crisis episodes. According to his research, it was meaningful to know the stock market crisis to predict the price, even though his research was not specific to certain prediction methods[48].

Abe, M. proposed a DNN network and his results showed that DNNs generally outperform shallow neural networks, and the best networks also outperformed representative machine learning models[49].

Nakagawa, K. proposed a deep factor model and a shallow model with DNN. The deep factor model outperformed the linear model. This implied that the relationship between the stock returns in the financial market and the factors is nonlinear, rather than linear. The deep factor model also outperformed other machine learning methods including SVR and random forest. The shallow model was superior in accuracy, while the deep model was more profitable[50].

Chong, E. proposed DNN networks and examined the effects of three unsupervised feature extraction methods including principal component analysis, auto-encoder, and the restricted Boltzmann machine, on the network's overall ability to predict future market behaviour. Empirical results suggested that DNNs could extract additional information from the residuals of the auto-regressive model and improve prediction performance; the same could not be said when the auto-regressive model is applied to the residuals of the network[51].

Singh, R. proposed a 2-Directional 2-Dimensional Principal Component Analysis (2D²PCA)+DNN which outperformed than the RNN and (2D²PCA) +RBFNN. The paper found that the best results were generated from a window size of 20 and a dimension of $10 \times 10$. The deep learning method for higher dimensions and large window sizes was giving a limited performance[52].

Yu, Pengfei, and Xuesong Yan proposed a novel DNN model which incorporated LSTM as well as phase-space recognition. The model had produce promising results. [87]

Yong, Bang Xiang, et al. propsed a DNN model with 40 nodes which showed a reasonable results and appear to be a highly profitable model. [89]

### 4.1.5 Reinforcement learning

8 articles that used reinforcement learning technology for the stock and Forex prediction were briefly described below, and as shown in the table. Table 6 showed the author, variables, dataset, and model of all the papers mentioned.

Table 6 The overall information for the papers which used Reinforcement learning model

| # | Author | Dataset | Variables | Model |
|---|--------|---------|-----------|-------|
| [57] | Li, Y. | US stock dataset | Close price and Volume | DQN |
| [58] | Shin, H.-G. | KOSPI | a candlestick chart, four moving average curves (5, 20, 60, and 120 days), , a bar graph of trading volume, DMI, and SSO | Reinforcement Learning combined with LSTM and CNN |
| [59] | Jia, W. | Chinese stock codes: 002415, 600016, 600028, 600547, 600999 and 601988 | Open price,high price, low price, close price, volume,DEA ,MACD EXPMA,CDP ,TRIX ,BBI, ASI ,KDJ ,RSI , PSY VR ,ADX ,CCI ,WR ,dm up,dm down | Reinforcement Learning with LSTM-based agent |
| [60] | Carapuço, J. | EUR/USD | bid/ask prices and volumes | Reinforcement Learning |
| [61] | Kang, Q. | S&P500 index | open, low, high, close price and trading volume | Reinforcement Learning with A3C algorithm |
| [62] | Zhu, Y. | S&P500 index | open, low, high, close price ,volume,MACD,MA6,M A12,RSI6,RSI12 and KD | Reinforcement Learning with ABN |
| [63] | Si, W. | Stock-IF,stock-IC,stock-IF | Close price | multi-objective deep reinforcement learning with LSTM |
| [64] | Pastore, A. | FTSE100 stock index | Date, Type, Stock, Volume, Price, and Total | Reinforcement learning |

Li, Y. proposed three different reinforcement learning methods to predict the price of the stock. The results showed that the best-performing Deep Reinforcement Learning model is DQN, not Double DQN. The paper also demonstrated Duelling DQN, which was an improved model based on DQN[57].

Shin, H.-G. proposed a Reinforcement Learning model that combined with LSTM and CNN. The model generated various charts from stock trading data and used them as inputs to the CNN layer. The features extracted through the CNN layer were divided into column vectors and inputted to the LSTM layer. The reinforcement learning defines agents' policy neural network structure, reward, and action, and provides buying, selling, and holding probabilities as final output[58].

Jia, W. proposed a Reinforcement Learning with an LSTM-based agent who could automatically sense the dynamics of the stock market and alleviate the difficulty of manually designing indicators from massive data. The paper had compared a wide range of different input sets[59].

Carapuço, J. proposed a reinforcement learning-Q network model, Three hidden layers of ReLU neutrons were trained as RL agents through the Q-learning algorithm under a novel simulated market environment framework. The framework was able to consistently induce stable learning that generalised to out-of-sample data[60].

Kang, Q. proposed to apply the state-of-art Asynchronous Advantage Actor-Critic algorithm(A3C algorithm) to solve the portfolio management problem and designed a standalone deep reinforcement learning model[61].

Zhu, Y. proposed an adaptive box-normalisation (ABN) stock trading strategy based on reinforcement learning, which improved the original box theory. In his ABN strategy, the stock market data was independently normalised inside each oscillation box[62].

Si, W. proposed a reinforcement learning model which had multi-objective and LSTM agent. It could be found that feature learning could contribute to better performances. The LSTM network made continuous decisions and could change positions at the right time, which reduced the transaction cost, and the multi-objective structure made good profits within the acceptable risk[63].

Pastore, A. analysed data for 46 players extracted from a financial market online game and tested whether Reinforcement Learning could capture these players' behaviour using a riskiness measure based on financial modelling[64].

### 4.1.6 Other Deep Learning Methods

8 articles that used other deep learning technology for the stock and Forex prediction were briefly described below, and as shown in the table. Table 7 showed the author, variables, dataset, and model of all the papers mentioned.

Table 7 The overall information for the papers which used other deep learning methods

| # | Author | Dataset | Variables | Model |
|---|--------|---------|-----------|-------|
| [65] | Long, W. | CSI300 | Open price, high price, low price, close price, Volume | MNFF |
| [66] | Wu, J.-L. | ANUE | stock message as information to form the text feature of each stock news(title, summary, and keywords) | HAN |
| [67] | Cho, C.-H. | CATHAY HOLDINGS, Fubon Financial, CTBC HOLDINGS, ESFH, and FFHC | Open price,high price, low price, close price, volume,MACD,CCI,ATR,BOLL,EMA12/20, MA5,MA1MOM6, MOM12,ROC,RSI, WVAD,Exchange rate and Interest rate | Wavenet |
| [68] | Minh, D.L | S&P 500, VN-index and cophieu68; Bloomberg, Reuters | Open price, high price, low price, close price, Volume, Stochastic oscillator, William (%R), and RSI Processed news article | Documents preprocessing-Documents labeling -Stock2Vec embedding-BGRU |
| [69] | Hu, G. | Financial Times Stock; Exchange 100 Index (FTSE 100) | Candlestick charts(images rather than annotation data) | Convolutional Auto-Encoder(CAE) |
| [70] | Hu, Z. | Chinese stock price; News(Not given specific data) | 1.Close price and volume; 2.news corpus sequence | HAN with SPL |
| [72] | Kim, T. and Khushi | Nine Dow Jones companies representing each sector - industrials (MMM), financials (JPM), | Open price, high price, low price, close price, Volume | 2D Relative-Attentional Gated Transformer |
| | | consumer services (PG), technology (AAPL), health care (UNH), consumer goods (WMT), oil & gas (XOM), basic materials (DD) and telecommunications (VZ) From Yahoo. | | |
| [74] | Zhang, and Khushi | Forex Exchange rates | Trend indicators: Moving average, Exponential moving average, Double exponential moving average, Triple exponential moving average, Vortex indicators Momentum indicators: Relative strength index, Stochastic oscillators, Volatility indicators: Bollinger bands, Ichimoku indicators | Genetic Algorithm |
| [79] | Hu, Ziniu | Chinese stiock data and economic news | News and financial data | Hybrid attention networks |
| [86] | Shi, Lei, et al. | Apple Inc. and S&P 500 | News and financial data | Hybrid of RNN, LSTM and CNN |

Long, W. proposed a novel end-to-end model named multi-filters neural network (MFNN) specifically for feature extraction on financial time series samples and price movement prediction task. Both convolutional and recurrent neurons are integrated to build the multi-filters structure. And the results showed that the proposed model is better than the RNN, CNN, and traditional machine learning methods[65].

Wu, J.-L. proposed a keyword-based attention network into Hierarchical Attention Networks (HAN), namely the HKAN model, to learn the relationships between dimensional sentiments (trend and trading) and stock messages which outperformed the HAN network[66].

Cho, C.-H.proposed three different models to predict the price of the stock, they were: LSTM，Seq2seq and Wavenet. According to his experiments, Wavenet outperformed the other two models[67].

Minh, D.L proposed a network to predict the directions of stock prices by using both financial news and sentiment dictionary. His results showed that TGRU achieved better performance than GRU and LSTM and Stock2Vec is more efficient in dealing with financial datasets[68].

Hu, G. proposed a Convolutional AutoEncoder model to learn a stock representation and converted a 4-channel stock time series (lowest, highest, open, and close price for the day) to candlestick charts by using synthesis technique to present price history as images. This method had successfully avoided expensive annotation. And the proposed model outperformed the FTSE 100 index and many well-known funds in terms of total return[69].

Hu, G. proposed a Hybrid Attention Networks (HAN) to predict the stock trend based on the sequence of recent related news. Moreover, he applied the self-paced learning(SPL) mechanism to achieve effective and efficient learning. The results showed the proposed model outperforms than RNN, Temporal- Attention-RNN, News-Attention-RNN, and HAN[70].

Kim, T. proposed a 2D relative-attentional gated transformer which was used to optimise portfolio return. It used general reinforcement learning with the agent incorporating 2D Relative-attentional Gated Transformer. [72]

Zhang, Z. proposed a genetic algorithm using a crossover of technique indicators as input. It had successfully outperformed some of the traditional trading strategies.[74]

Hu, Ziniu proposed hybrid attention networks which is a novel way of combining financial time-series information and news nature language information. [79]

Shi, Lei, et al. propsed a system where it perform factor analysis then utilise multiple depp learning method to design a model which outperformed benmark. [86]

### 4.2 Papers Results grouped by the method used

#### 4.2.1 CNN

Since the performance metrics used in different articles were different, our survey analysed them based on various metrics. Results used were the average performance of the best performing models in the mentioned papers. Table 8 showed the metrics and results for the papers which used the CNN model.

Table 8 The metrics and results for the papers which used the CNN model

| Performance Metrics | Reference no. | Corresponding value | Performance Metrics | Reference no. | Corresponding value |
|---|---|---|---|---|---|
| RMSE | [8] | 0.043+/- .007 | MAPE | [9] | 5 |
| | [9] | 11 | Sharpe ratio | [11] | [11]2D:0.1422, 3D:0.1413 |
| | [71] | 0.395185*10^-3 | | | |
| MAE | [9] | 6 | | [15] | [15]0.611 |
| | [71] | 0.240977*10^-3 | | | |
| Accuracy | [10] | 71% | CEQ | [11] | 2D:0.0006681 3D:0.000664 |
| | [13] | 60.02% | Return rate | [11] | [11](2D:1.2312 |
| | [16] | 55.44% | | | |
| | [19] | 71.72% | | | 3D:1.2604) |
| | [20] | 65.08% | | [13] | [13](1.3107) |
| | [76] | 75.2% | | | |
| | [77] | 78.46% | | | |
| | [78] | 57.88% | | | |
| | [90] | 74.753% | | | |
| Error Percentage | [17] | 95.02% | | [15] | [15](1.2156) |
| F-measure | [11] | 2D:0.4944 | | [18] | [18](1.309) |
| | | 3D:0.4931 | Mean Test Score | [12] | 0.000281317 |
| | [15] | 0.6227 | MSE | [19] | 0.2631 |
| | | | | [71] | 0.156*10^-6 |
| | [16] | 0.7133 | AE | [8] | 0.029+/-.005 |
| | [76] | 0.73 | | | |
| | [90] | 0.6367 | | | |

#### 4.2.2 RNN

Since the performance metrics used in different articles were different, our survey analysed them based on different metrics. Results used were the average performance of the best performing models in the mentioned papers. Table 9 shows the metrics and results for the papers which used the RNN model.

Table 9 The metrics and results for the papers which used RNN model

| Performance Metrics | Reference no. | Corresponding value |
|---|---|---|
| RMSE | [53] | 512-530 |
| | [55] | 0.0205 |
| | [73] | 0.00165 |
| MAPE | [55] | 0.2431 |
| | [73] | 0.232 |
| MAE | [55] | 0.0132 |
| Accuracy | [54] | 68.95% |
| | [55] | 66.54% |
| F-measure | [54] | 0.7658 |
| Recall | [54] | 0.7471 |
| Precision | [54] | 0.7855 |
| MSE | [56] | 0.057443 |

#### 4.2.3 LSTM

Since the performance metrics used were different in different articles, our survey would analyse them based on different metrics. Results used were the average performance of the best performing models in the mentioned papers. Table 10 showed the metrics and results for the papers which used the LSTM model.

Table 10 The metrics and results for the papers which used LSTM model

| Performance Metrics | Reference no. | Corresponding value | Performance Metrics | Reference no. | Corresponding value |
|---|---|---|---|---|---|
| RMSE | [21] | 0.306543 | MAPE | [24] | 1.03 |
| | [24] | 347.46 | | [27] | 1.6905 |
| | [25] | 0.0151 | | [29] | 4.13 |
| | [26] | 25.90 | | [30] | 1.0077 |
| | | | | [75] | 0.119 |
| | | | | [82] | 0.91 |
| | | | | [84] | 1.37 |
| | | | | [91] | 1.65 |
| | | | | [95] | 0.6346 |
| | [27] | 0.0242 | Precision | [42] | 0.553 |
| | [28] | 1.3 | Recall | [42] | 0.129 |
| | [34] | 9.72 | Return rate | [22] | 1.0667 |
| | [35] | 4.24(Average) | MSE | [21] | 0.093969 |
| | [36] | 1-10 | | [22] | 0.004845492 |
| | [75] | 0.0015 | | | |
| | [94] | 0.02295 | | [81] | 0.000379 |
| MAE | [21] | 0.21035 | | [24] | 120731.4 |
| | [24] | 262.42 | | [27] | 19.7096 |
| | [26] | 0.1895 | | [28] | 0.019 |
| | [27] | 0.0169 | | [29] | 0.00098 |
| | [29] | 0.023 | | [30] | 7.56 |
| | [31] | 0.01069 | | [31] | 0.00149 |
| | [30] | 1.975 | | [32] | 1.012 |
| Accuracy | [23] | 54.58% | MCC | [23] | 0.0478 |
| | [25] | 98.49% | R2 | [24] | 0.83 |
| | [34] | 60.60% | HMAE | [31] | 0.42911 |
| | [39] | 65% | HMSE | [31] | 0.23492 |
| | [41] | 83.91% | IC | [37] | 0.1259 |
| | [42] | 55.90% | AR | [37] | 0.2015 |
| | [43] | 63.34% | IR | [37] | 3.0521 |
| | [45] | 27.20% | | | |
| | [80] | 53.2% | | | |
| | [83] | 87.86% | | | |
| | [91] | 70.56% | Score | [44] | 0.4271 |
| | [92] | 75.89% | | | |
| | [93] | 75.58% | | | |
| F-measure | [42] | 0.209 | | | |

## 4.2.4 DNN

Since the performance metrics used in different articles were different, our survey analysed them based on different metrics. Results used were the average performance of the best performing models in the mentioned papers. Table 11 showed the metrics and results for the papers which used the DNN model.

Table 11 The metrics and results for the papers which used DNN model

| Performance Metrics | Reference no. | Corresponding value | Performance Metrics | Reference no. | Corresponding value |
|---|---|---|---|---|---|
| RMSE | [50] | 0.0951 | Sharpe ratio | [50] | 1.41 |
| | | | | [89] | 5.34 |
| | [51] | 0.8214 | Return rate | [49] | 1.0952 |
| | [52] | 0.00674 | | [50] | 1.1081 |
| MAE | [50] | 0.0663 | CORR | [49] | 0.0582 |
| | [51] | 0.5852 | MSE | [49] | 0.0836 |
| Accuracy | [46] | 61.90% | | [51] | 0.9621 |
| | [47] | 84.50% | SMAPE | [52] | 0.0696 |
| | [87] | 58.07% | | | |

| | | | MAPE | [52] | 0.080059 |
|---|---|---|---|---|---|
| F-measure | [47] | 0.824 | | [89] | 1.84 |
| | | | Volatility | [50] | 7.65% |

#### 4.2.5 Reinforcement learning

Since the performance metrics used were different in different articles, our survey would analyse them based on different metrics. Results used were the average performance of the best performing models in the mentioned papers. Table 12 showed the metrics and results for the papers which used the Reinforcement learning model.

Table 12 The metrics and results for the papers which used Reinforcement learning model

| Performance Metrics | Reference no. | Corresponding value |
|---|---|---|
| Sharpe ratio | [58] | 2.77 |
| | [63] | 0.12 |
| Return rate | [59] | 1.948 |
| | [60] | 1.163 $\pm$ 2.8% |
| | [62] | 2.442 |
| MSE | [62] | 0.000412 |

#### 4.2.6 Other Deep Learning Methods

Since the performance metrics used in different articles were different, our survey analysed them based on different metrics that were used. Results used were the average performance of the best performing models in the mentioned papers. Table 13 showed the metrics and results for the papers which used other deep learning methods.

Table 13 The metrics and results for the papers which used other deep learning methods

| Performance Metrics | Reference no. | Corresponding value |
|---|---|---|
| RMSE | [67] | 0.6866 |
| Accuracy | [68] | 66.32% |
| | [70] | 47.8% |
| | [86] | 79.7% |
| Sharpe ratio | [65] | 4.49. |
| | [69] | 0.8 |
| | [72] | 0.6418 |
| | [74] | 6.68 on EURUSD currency |
| Return rate | [65] | 1.4228 |
| | [68] | 1.0531(0.25%) |
| | [69] | 1.118 |
| | [70] | 1.611(0.3%) |
| | [72] | 1.4316 |
| | [74] | 1.0968 on EURUSD currency |
| | [79] | 1.52 |
| MSE | [66] | 1.05 |
| MDAE | [66] | 0.71 |
| Correlation | [67] | 0.9564 |
| Precision | [68] | 72.1% |
| Recall | [68] | 77.32% |

## 5. Discussion

### 5.1 Analysis based on the method used

#### 5.1.1 CNN

Some findings can be drawn from reviewing the CNN models:

1. According to the datasets that all the paper used, 6 papers used the combination of technical analysis and sentiment and news analysis to predict the stock. The rest of them used the method of technical analysis only.

2. As for the variables, the close price was the choice of all CNN models, and five papers were using close price only.

3. It could be found that 12 of the papers changed the traditional CNN model to pursue higher performance of prediction. And the combination of CNN and LSTM was the most common model.

4. The metrics used in each paper were different; there were 11 metrics used for measuring the performance of the CNN model.

5.Multiple articles selected RMSE, Return rate, F-measure, Sharpe ratio, and accuracy. We could find that the paper [19] had the highest accuracy than any other papers, paper [13] had the highest return rate followed by the paper [18], [11] and [15]. Paper [71] had the lowest RMSE. Paper [16] had a higher F-measure than paper [11], and [15], but paper [15] achieved a higher Sharpe ratio than paper [11].

### 5.1.2 RNN

In the section of the RNN based model, the following conclusions could be drawn:

1. According to the datasets that all the paper used, there was 1 paper that used the combination of technical analysis and sentiment and news analysis to predict the stock. And the rest of them used technical analysis only.

2. For the variables, all the RNN based models used the multivariable input and open price, close price, highest price, and the lowest price are used in all models as an input.

3. It could be found that all the papers changed the traditional RNN model to pursue higher performance of prediction. Two of the papers chose the C-RNN based model.

4. The metrics used in each paper were different; in total, there were 8 metrics used for measuring the performance of the RNN based model.

5.RMSE and accuracy were selected by multiple articles. We could note that paper [55] has a much lower RMSE than paper [53], and paper [54] had higher accuracy than paper [55].

### 5.1.3 LSTM

Following points were worth discussing from reviewing the LSTM papers:

1. According to the datasets that all the paper used, 3 papers used a combination of technical analysis and sentiment and news analysis to predict the stock. The rest of them used technical analysis only with the exception of one paper which used expert recommendations.

2. As for the variables, the close price was the choice of 23 LSTM based models, there were 8 papers using close price only, and 12 papers selected to include close price, open price, high price, and low price in their input.

3. It could be found that 15 of the papers changed the traditional LSTM model to pursue higher performance of prediction. Attention-based LSTM and LSTM with RNN were the most frequent models that showed up in three different papers. Two papers chose the method of LSTM with GRU to improve the model.

4.The metrics used in each paper were different; there were 17 metrics used for measuring the performance of the LSTM based model.

Multiple articles selected 5.RMSE, MAPE, MAE, accuracy, and MSE, we could find that Paper [25], paper [94] and paper [27] was in the lowest order of magnitude with paper [25] achieving the lowest RMSE; paper [21] was in the second-lowest order of magnitude; paper [28],[35],[36] and paper [34] were in the third lowest order of magnitude. Paper [26] was in the fourth lowest order of magnitude and paper [24] had the highest order of magnitude.

6. As for MAPE, paper [75] had the lowest order of magnitude followed by the paper [30] ,[24], [82], [84],[91],[95] and [27]. And paper [29] had the highest order of magnitude of MAPE.

7. As for MAE, paper [75], [31], [27] and [29] had the lowest order of magnitude MAE with paper [75] having the best performance; paper[26] and [21] were in the second-lowest order of magnitude while paper [30] was in the third lowest order of magnitude. Paper [24] had the highest order of magnitude.

8. As for MSE, paper [81]had the lowest MSE and was in the lowest order of magnitude; paper [29] , [31] and [22] were in the second-lowest order of magnitude while paper [28] and [21] were in the third lowest order of magnitude. Paper [32] and [30] were in the fourth lowest order of magnitude, paper [27] was in the fifth-lowest order of magnitude, and paper [24] had the highest order of magnitude of MAE.

9.As for accuracy, paper [25] had the highest accuracy followed by [83], [41],[39], [43], [34], [42], [23] and paper [45] had the lowest accuracy.

### 5.1.4 DNN

In the section of DNN-based model, the following conclusions could be drawn:

1. According to the datasets that all the paper used, no paper used the combination of technical analysis and sentiment and news analysis to predict the stock. All of them used the method of technical analysis only.

2. As for the variables, six of the seven DNN-based models used multivariable input, and only one paper used close price as its sole input.

3. It could be found that 3 of the papers changed the traditional DNN model to pursue higher performance of prediction. All of the improved models were not duplicated.

4. Because the metrics used in each paper were different; there were 11 different metrics used for measuring the performance of the DNN-based model.

5. Multiple articles selected 5.RMSE, MAE, accuracy, return rate, and MSE. We could find that the paper [52] had the lowest RMSE followed by the paper [50] and [51], with the latter two in different orders of magnitude.

6. As for MAE, paper [50] had a lower MAE than paper [51], and the accuracy of the paper [47] was higher than paper [46] and [87]. Furthermore, paper [50] had a higher return rate than paper [49]. And paper [49] had a lower MSE than paper [51].

### 5.1.5 Reinforcement learning

In the section of Reinforcement learning-based model, the following conclusions could be drawn:

1. According to the datasets that all the paper used, no paper used the combination of technical analysis and sentiment and news analysis to predict the stock. All of them used technical analysis only.

2. As for the variables, 7 of the reinforcement learning-based models used the multivariable input, and only one paper solely used close price as input.

3. It could be found that three of 6 papers changed the traditional Reinforcement learning model to pursue higher performance of prediction. 3 of the improved models were combined with LSTM.

4. The metrics used in each paper were different; in total, there were 3 metrics used in the measurement performance of the Reinforcement learning-based model.

5. Multiple articles selected 5. Sharpe ratio and return rate, we could find that paper [58] had a much higher Sharpe ratio than paper [63]; paper [62] had a higher return rate than paper [59] then followed by the paper[60].

### 5.1.6 Other Deep Learning Methods

In the section of other deep learning methods based model, the following conclusions could be drawn:

1. According to the datasets that all the paper used, 4 papers made use of sentiment and news analysis to predict the stock. The rest of them used the method of technical analysis only.

2. As for the variables, five of the other deep learning methods models used the multivariable input and only one paper used candlestick charts alone as input.

3. It could be found that there were 5 different models in the other deep learning methods section. The only model that appeared three times was HAN, which consist of an ordinary HAN model and two modified HAN models. The rest of the models were not duplicated.

4. Because the metrics used in each paper were different, there were 9 metrics used in the measurement performance of this section.

5. Multiple articles selected 5. Accuracy, Sharpe ratio, and Return rate, we could find that the paper [86] had the highest accuracy.Paper [65] had a much higher Sharpe ratio than paper [69].

6. As for the return rate, paper [70] had the highest return rate followed by the paper [65], [69]and paper [68] had the lowest return rate.

*5.2 Discussion and Analysis based on performance metrics*

In this part, all of Forex/Stock price prediction models mentioned above, which made use of the deep learning, would be discussed and analysed. All analysis of the performance metric data was shown in Table 14. It could be found that the most commonly used performance metrics were RMSE, MAPE, MAE, MSE, Accuracy, Sharpe ratio, and Return rate. Hence our review would analyse the papers in terms of its performance metrics used.

Table 14 Analysis based on Performance Metrics

| Performance Metrics | Reference no. |
|---|---|
| RMSE | [8], [9], [21], [24], [25], [26], [27], [28], [34], [35],[36],[50], [51], [52], [53], [55], [67],[71],[73] |
| MAPE | [9], [24], [27], [29], [30], [52], [55],[73],[82],[84],[94] |
| MAE | [9], [21], [24], [26], [27], [29], [31], [30], [50], [51], [55],[71],[95] |
| Accuracy | [10],[13],[16], [19], [20],[23], [25], [34],[39], [41], [42], [43], [45],[46], [47], [54], [55], [68], [70],[76],[77],[78],[80],[83],[87],[90],[91],[92],[93] |
| F-measure | [11], [42], [47], [54],[15],[16],[76],[90] |
| Sharpe ratio | [11], [15],[50],[58], [63], [65], [69], [72],[74] |
| CEQ | [11] |
| Return rate | [11], [13],[15],[18], [22], [49], [50],[59], [60], [65],[68],[69],[70] , [72],[74],[79] |
| Mean Test Score | [12] |
| MSE | [19], [21], [22], [24], [27], [28], [29], [30], [31], [32], [49], [51],[56],[62],[71],[81] |
| AE | [8] |
| Precision | [42], [54] |
| Recall | [42], [54] |
| R2 | [24] |
| Error Percentage | [17] |
| MCC | [23] |
| HMAE | [31] |
| HMSE | [31] |
| CORR | [49] |
| SMAPE | [52] |
| Volatility | [50] |
| IC | [37] |
| AR | [37] |
| IR | [37] |
| Score | [44] |

5.2.1 analysis based on RMSE

In this part, the value of the RMSE in the papers would be discussed. Table 15 showed the details of each paper's RMSE based on the numerical range. In this table, paper [8], [9], [71] used CNN based model; paper   [21], [24], [25], [26], [27], [28], [34],[35],[36], [75],[94] used LSTM based model; paper [50], [51], [52] used DNN based model; paper [53], [55], [73] used RNN based model and paper [67] used other deep learning method.

Table 15 Analysis based on RMSE

| RMSE range | Reference no. |
|---|---|
| RMSE | [8], [9], [21], [24], [25], [26], [27], [28], [34], [35],[36],[50], [51], [52], [53], [55], [67],[71],[75],[94] |
| <0.001 | [71], [75] |
| 0.001-0.01 | [52], [73] |
| 0.01-0.1 | [8], [21], [25],[27],[50], [55],[94] |
| 0.1-1 | [9],[51],[67] |
| 1-10 | [28],[34],[35],[36] |
| 10-100 | [26] |
| >100 | [24], [53] |

It was clear that paper [71], [75] achieved the best performance using DNN model. Papers that had a RMSE smaller than 0.001, paper [52], [73], [8], [21], [25], [27], [50], [55] had great performance while paper [26], [24], [53] didn't have the low RMSE .

Among all the papers, 33% of the CNN papers had an RMSE below 0.001, 33% of the CNN papers were in the range of 0.01-0.1, and 33% of the CNN papers were in the range of 0.1-1. 36% of the LSTM papers were in the range of 0.01-0.1 and 1-10 respectively, , the rest LSTM papers were distributed equally in range of <0.001, 10-100 and above 100 with 9%in each range. 33.3% of the DNN papers were in the range of 0.001-0.01, 0.01-0.1 and 0.1-1 respectively. The RNN papers were distributed equally in the range of 0.001-0.01, 0.01-0.1 and above 100, i.e. 33% in each range. The only paper that used other deep learning method had an RMSE in the range of 0.01-0.1.

### 5.2.2 analysis based on MAPE

In this part, the value of the MAPE would be discussed. Table 16 showed the papers' MAPE performance based on the numerical range. In this table, paper [9] used CNN based model; paper [24], [27], [29], [30], [75], [82], [84], [95] used LSTM based model; paper [52],[89] used DNN based model; paper [55] used RNN based model.

Table 16 Analysis based on MAPE

| MAPE range | Reference no. |
|---|---|
| MAPE | [9], [24], [27], [29], [30], [52], [55],[75],[82],[84],[89], [95] |
| 0-0.5 | [52],[55],[75], [95] |
| 0.5-1 | [82] |
| 1-1.5 | [24],[30],[84],[89] |
| 1.5-2 | [27] |
| 2-10 | [9],[29] |

It could be found that paper [52], [55],[75] performed best which used the DNN model and RNN model and LSTM respectively, and both were in the range of 1-1.5, paper [24],[30],[82] had great performance while paper [9], [29] didn't have a low MAPE.

Among all the papers, the only CNN paper was in the range of 2-10. 37.5% of the LSTM papers were in the range of 1-1.5, 12.5% of the LSTM papers were in the range of 0.5-1, 1.5-2, and 2-10 respectively, 25% of LSTM papers in a range of 0-0.5. The two DNN papers were in the range 0-0.5, and 1-1.5. Lastly, the only one RNN paper was in the range of 0-0.5.

### 5.2.3 analysis based on MAE

In this part, the value of the MAE would be discussed. Table 17 showed the details of papers' MAE. In this table, paper [9], [71] used CNN based model; paper [23], [26], [28], [29], [31], [33], [32] used LSTM based model; paper [50], [51] used DNN based model and paper [55] , [73] used RNN based model.

Table 17 Analysis based on MAE

| MAE range | Reference no. |
|---|---|
| MAE | [9], [21], [24], [26], [27], [29], [31], [30], [50], [51], [55], [71] |
| <0.01 | [71] |
| 0.01-0.1 | [27],[29],[31],[50],[55] |
| 0.1-1 | [21],[26],[51],[73] |
| 1-10 | [9],[30] |
| 10-100 | N/A |
| >100 | [24], |

It could be found that paper [71] performed best which used CNN model. Papers' MAE were in the range of 0.1-1. Paper [27], [29], [31], [50], [55] also had good performance, however, paper [24] didn't have a low MAE.

Among all the papers that used MAE as measurement, 50% of CNN paper was in the range of 1-10; the other 50% had an MAE small than 0.01. 42.9% of the LSTM papers were in the range of 0.01-0.1, 28.6% of the LSTM papers were in the range of 0.1-1, 14.2% of the LSTM papers were in the range of 1-10 and above 100 respectively, 50% of the DNN papers were in the range of 0.01-0.1 and 0.1-1 respectively. The two RNN papers were in the range of 0.01-0.1 and 0.1-1.

### 5.2.4 analysis based on MSE

In this part, the value of the MSE would be discussed. Table 18 showed the details of papers' MSE. In this table, paper [19], [71] used CNN based model; paper [21], [22], [24], [27], [28], [29], [30], [31], [32], [81] used LSTM based model and paper [49], [51] used DNN based model. Paper,[56] used RNN based model, and paper [62] used the Reinforcement learning based model.

Table 18 Analysis based on MSE

| MSE range | Reference no. |
|---|---|
| MSE | [19], [21], [22], [24], [27], [28], [29], [30], [31], [32], [49], [51],[56],[62],[71] |
| <0.01 | [71],[81] |
| 0-0.01 | [22],[29],[31],[62] |
| 0.01-0.1 | [21],[28],[49],[56] |
| 0.1-1 | [19],[32],[51] |
| 1-10 | [30] |
| 10-100 | [27] |
| >100 | [24] |

It could be found that paper [81] performed the best which used the LSTM model; its MSE was smaller than 0.01. Paper [71], [22], [29], [31],[62] also had good performance, however, paper [24], [27] didn't have the low MSE.

Among all the papers that used MSE as performance measurement, the only CNN paper was in the range of 0.1-1, 40% of the LSTM papers were in the range of 0-0.01, 20%of the LSTM papers were in the range of 0.01-0.1, 10% of the LSTM papers were in the range of 0.1-1, 1-10, 10-100 and above 100 respectively, 50% of the DNN papers were in the range of 0.01-0.1 and 0.1-1 respectively. The only RNN based model and reinforcement learning-based model was in the range of 0-0.01 and 0.01-0.1 respectively.

### 5.2.4 analysis based on the accuracy

In this part, the value of the accuracy would be discussed. Table 19 shows the details of papers' accuracy. In this table, paper [10], [13],[16], [19],[20],[76],[77],[78],[90] used CNN based model; paper [23], [25], [34], [39],[41], [42], [43],[45], [80], [83], [91],[92],[93]

used LSTM based model;paper [46], [47], [87] used DNN based model; paper [54], [55] used RNN based model and paper [68], [70],[86] used other deep learning method.

Table 19 Analysis based on the accuracy

| Accuracy range | Reference no. |
|---|---|
| Accuracy | [10],[13],[16], [19], [20],[23], [25], [34],[39], [41], [42], [43], [45],[46], [47], [54], [55], [68], [70],[76],[77],[78],[80],[83],[86],[87], [90],[91],[92],[93] |
| 0-50% | [45],[70] |
| 50%-60% | [16],[23],[42],[78],[80],[87] |
| 60%-70% | [13], [20],[34],[39],[43],[46],[54], [55],[68] |
| 70%-80% | [10],[19],[76],[77],[86],[90],[91],[92],[93] |
| 80%-90% | [41],[47],[83] |
| 90%-100% | [25] |

It could be found that paper [25] performed best, which used LSTM model, and the paper's accuracy was in the range of 90%-100%. Paper [41], [47] ,[83] also had great performance, but paper [45], [70] didn't have the high accuracy.

Among all the papers that had accuracy measure metric, 22% of the CNN papers were in the range of 50%-60%, and 22% of the CNN papers were in the range of 60%-70% and 55% of the CNN papers were in the range of 70%-80%. 8% of the LSTM papers were in the range of 0-50%, and 90%-100% respectively,15% of the LSTM papers were in the range of 80%-90%, 23% of the LSTM papers were in the range of 50%-60%, 60%-70% and 70%-80% respectively. 33% of the DNN papers were in the range of 50%-60%, 60%-70% and 80%-90% respectively. All the RNN papers were in the range of 60%-70%. 33% of other deep learning methods were in the range of 40%-50%, 60%-70% and 70-80% respectively.

### 5.2.5 analysis based on Sharpe ratio

In this part, the value of the Sharpe ratio would be discussed. Sharpe ratio referred to the amount of return generated relative to the risk taken, where risk was calculated using the standard deviation of return. Table 20 showed the details of the papers' that used the Sharpe ratio as a preformance metric. In this table, paper [11] and [15] used CNN based model; paper [50], [89] used DNN based model; paper [58], [63] used Reinforcement learning-based model, and paper [65], [69], [72], [74] used other deep learning method.

Table 20 Analysis based on Sharpe ratio

| Sharpe ratio range | Reference no. |
|---|---|
| Sharpe ratio | [11], [15],[50],[58], [63], [65], [69],[72],[74] |
| 0.1-1 | [11], [15],[63],[69],[72] |
| 1-2 | [50] |
| 2-5 | [58] |
| 5-10 | [65],[74],[89] |

It could be found that paper [65], [74],[89] performed best, which used DNN model, and Sharpe ratio was in the range of 5-10. Paper [58] also had a great performance, but paper [11], [15],[63], [69], [72] didn't have the high Sharpe ratio.

Among all the paper that used the Sharpe ratio, all CNN papers were in the range of 0.1-1. The two DNN papers were in the range of 1-2 and 5-10. 50% of the Reinforcement learning paper was in the range of 0.1-1, and 50% of the Reinforcement learning paper was in the range of 2-5. 50% of the other deep learning method paper was in the range of 0.1-1, and 50% of the other deep learning method paper was in the range of 5-10.

### 5.2.6 analysis based on the Return rate

In this part, the value of the return rate would be discussed. Table 21 showed the details of papers' return rate. In this table, paper [11],[13],[15], [18] used CNN based model; paper [22] used LSTM based model; paper [49], [50] used DNN based model; paper [59], [60] used Reinforcement learning based model and paper [65], [68], [69], [70], [72], [74], [79] used other deep learning method.

Table 21 Analysis based on the return rate

| Return rate range | Reference no. |
|---|---|
| Return rate | [11], [13],[15],[18], [22], [49], [50],[59], [60], [65],[68],[69],[70] |
| 1.0-1.2 | [22], [49], [50],[60],[68],[69], [74] |
| 1.2-1.4 | [11],[13],[15],[18] |
| 1.4-1.6 | [65],[72],[79] |
| 1.6-1.8 | [70] |
| 1.8-2.0 | [59] |

It could be found that paper [59] performed the best which used Reinforcement learning model, and the paper's return rate was in the range of 1.4-1.8. Paper [65] and [70] had good performance but paper [22], [49], [50], [60], [68], [69], [74] didn't have the high return rate.

Among all the papers that used return rate as a performance measurement, all the CNN papers were in the range of 1.2-1.4, and the only LSTM paper was in the range of 1.0-1.2, all the DNN papers were in the range of 1.0-1.2. 50% of the Reinforcement learning papers were in the range of 1.0-1.2, and 50% of the Reinforcement learning papers were in the range of 1.8-2. The other deep learning method papers performed well with 14% paper in the range of 1.0-1.2, 42% paper in the range of 1.4-1.6 , and 42% paper in the range of 1.6-1.8.

## 6. Conclusions

This paper provided a detailed review of 88 papers from 2015 to the present on predicting stock /Forex price movements through deep learning methods. The existing stock/Forex models were evaluated through analysing data sets, variables, the use of different models, and the metrics of evaluation. The research review included a wide range of techniques: CNN, LSTM, DNN, RNN, Reinforcement learning, and other deep learning methods such as HAN, NLP, and Wavenet. Furthermore, the data sets, variables, models, and their different results were analysed and compared within each technique. Then our paper discusses the main performance metrics of all models. They are RMSE, MAPE, MAE, MSE, Accuracy, Sharpe ratio and Return rate.

This paper aimed to contribute to the research of stock/ Forex market prediction through the analysis of the above different deep learning prediction models. Through the review, It can be identified that there is a lack of studies on the combination of multiple deep learning methods, espically in respect to other deep learning methods. The hybrid networks are showing promising signs for future research. In the future, we would design a specific hybrid model based on the above analysis, incorporating latest technology such as advanced genetic algorithms, self attention neural networks to predict the stock/Forex market.

**Author Contributions:** First two authors contributed equally, MK conceived & designed the study supervised the work and revised the manuscript.

**Funding:** This research received no external funding.

**Conflicts of Interest:** The authors declare no conflict of interest.